\documentclass[]{spie}  

\usepackage{amsmath,amsfonts,amssymb}
\usepackage{graphicx, gridline}
\usepackage{multirow}
\graphicspath{ {./figures/} }
\usepackage[colorlinks=true, allcolors=blue]{hyperref}

\title{Liger at Keck Observatory: Imager Detector and IFS Pick-off Mirror Assembly}

\author[a,b]{Maren Cosens}
\author[a,b]{Shelley A. Wright}
\author[a]{Aaron Brown}
\author[c]{Michael Fitzgerald}
\author[c]{Chris Johnson}
\author[d]{Tucker Jones}
\author[e]{Marc Kassis}
\author[c]{Evan Kress}
\author[f]{Renate Kupke}
\author[c]{James E. Larkin}
\author[c]{Kenneth Magnone}
\author[e]{Rosalie McGurk}
\author[a]{Nils-Erik Rundquist}
\author[c]{Ji Man Sohn}
\author[c]{Eric Wang}
\author[a,b]{James Wiley}
\author[e]{Sherry Yeh}
\affil[a]{Center for Astrophysics \& Space Sciences, University of California San Diego, USA}
\affil[b]{Department of Physics, University of California San Diego, USA}
\affil[c]{Department of Physics \& Astronomy, University of California Los Angeles, USA}
\affil[d]{Department of Physics, University of California Davis, USA}
\affil[e]{W.M. Keck Observatory, Waimea, HI}
\affil[f]{Department of Astronomy \& Astrophysics, University of California Santa Cruz, USA}

\authorinfo{Further author information: (Send correspondence to M.C.)\\M.C.: E-mail: mcosens@ucsd.edu}

\begin{document} 
\maketitle

\begin{abstract} 
Liger is a next-generation near-infrared imager and integral field spectrograph (IFS) planned for the W.M. Keck Observatory. Liger is designed to take advantage of improved adaptive optics (AO) from the Keck All-Sky Precision Adaptive Optics (KAPA) upgrade currently underway. Liger operates at 0.84-2.45 $\mu$m with spectral resolving powers of R$\sim$4,000-10,000. Liger makes use of a sequential imager and spectrograph design allowing for simultaneous observations. There are two spectrograph modes: a lenslet with high spatial sampling of 14 and 31 mas, and a slicer with 75 and 150 mas sampling with an expanded field of view. Two pick-off mirrors near the imager detector direct light to these two IFS channels. We present the design and structural analysis for the imager detector and IFS pick-off mirror mounting assembly that will be used to align and maintain stability throughout its operation. A piezoelectric actuator will be used to step through $\rm3\,mm$ of travel during alignment of the instrument to determine the optimal focus for both the detector and pick-off mirrors which will be locked in place during normal operation. We will demonstrate that the design can withstand the required gravitational and shipping loads and can be aligned within the positioning tolerances for the optics.   
\end{abstract}


\section{INTRODUCTION}\label{sec:intro}
Liger is a new integral field spectrograph (IFS) and imager in development for the W.M. Keck Observatory which will take advantage of the ongoing Keck All-Sky Precision Adaptive Optics (KAPA) upgrade. Liger\cite{Wright2019, Wright2022, Cosens2020, Wiley2020} will provide a number of improvements over existing AO fed instruments including larger fields of view, finer spectral resolution (up to $\rm R\sim8,000 - 10,000$), and extending to bluer wavelengths ($\rm 0.84-2.45\,\mu m$). The Liger design draws from the heritage of two key sources: the imager component is custom designed for Liger but makes use of similar mechanisms to the Keck OSIRIS imager\cite{Larkin2006}, and the spectrograph is a clone of the design developed for the InfraRed Imaging Spectrograph (IRIS)\cite{Larkin2016, Larkin2020, Zhang2018} --- the planned first light instrument for the Thirty Meter Telescope. Like IRIS, Liger will have two spectrograph channels, a slicer and lenslet mode, which will share a common grating turret, three mirror anastigmat cameras, and detector. The Liger imager filter and pupil wheel mechanisms make use of similar gear, detent, and limit switch designs as OSIRIS, but with improvements to the number of filter slots and the presence of a dedicated pupil wheel at the pupil location\cite{Cosens2020}. For a full overview of Liger see Wright et al. (2019)\cite{Wright2019} and Wright et al. (\textit{this conference})\cite{Wright2022}.

Here we present the design of the assembly which will be used to mount and align the Liger imager detector as well as the pick-off mirrors which feed the two spectrograph modes. These two components require a common assembly to place the pick-off mirrors as close to the imager focal plane as possible for the best optical performance. Key requirements for the assembly are listed in Table \ref{tab:requirements}. 

\begin{table}[h]
    \centering
    \caption{Key Requirements: Detector and Pick-off Mirrors}
    \label{tab:requirements}
    \begin{tabular}{|l|l|}
    \hline
        \textbf{Parameter} &  \textbf{Value} \\
        \hline \hline
        Operating Temperature & $\rm 77\,K$ \\
        \hline
        Operating Pressure & $\rm 10^{-5}\,Torr$ \\
        \hline
        Shock Load & $\rm 4g$ ($+$ gravity) \\
        \hline
        Resonant Frequency to Avoid & $\rm8-80\,Hz$ \\
        \hline
        Focus Travel & $\rm3\,mm$ \\
        \hline
        Focus Offset Range & \multirow{2}{4em}{$\rm1\,mm$} \\
        (pick-offs) & \\
        \hline
        Focus Accuracy & $\rm100\,\mu m$ \\
        \hline
        Tip-tilt Range & \multirow{2}{4em}{$\rm2^\circ$} \\
        (detector) & \\
        \hline
        Tip-tilt Accuracy & \multirow{2}{4em}{$\rm0.25^\circ$} \\
        (detector) & \\
        \hline
        Tip-tilt Accuracy & \multirow{2}{4em}{$\rm0.2^\circ$} \\
        (pick-offs) & \\
    \hline
    \end{tabular}
\end{table}

The design of the detector-mirror assembly is outlined in Section \ref{sec:design} including the adjustability (Section \ref{sec:focus} \& \ref{sec:adjustment}) and baffling (Section \ref{sec:baffling}) necessary to achieve the instrument requirements. In Section \ref{sec:structural_analysis}, structural analysis is performed to verify performance of the assembly under the loads and frequencies specified in Table \ref{tab:requirements}. 

\section{MECHANISM DESIGN}\label{sec:design}
\begin{figure}[h]
    \centering
   \gridline{\fig{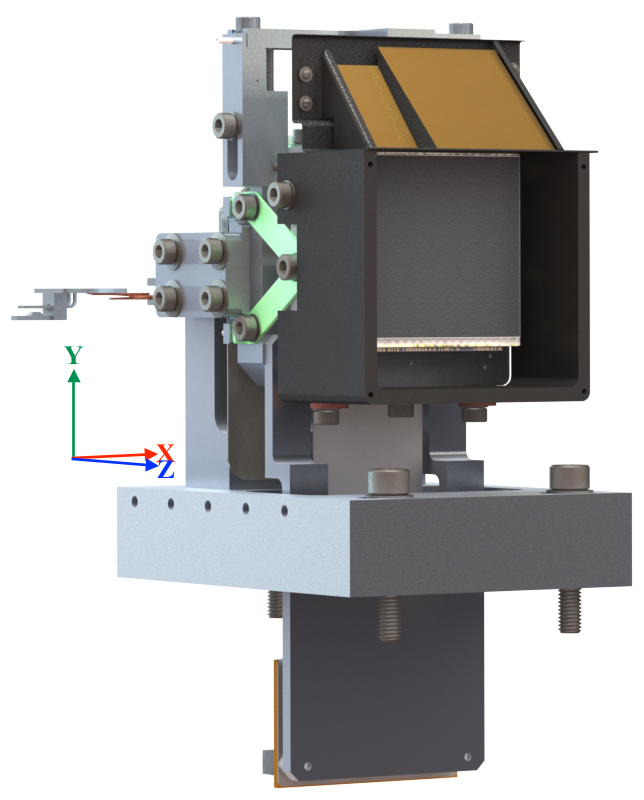}{0.48\textwidth}{(a): rendering}
            \fig{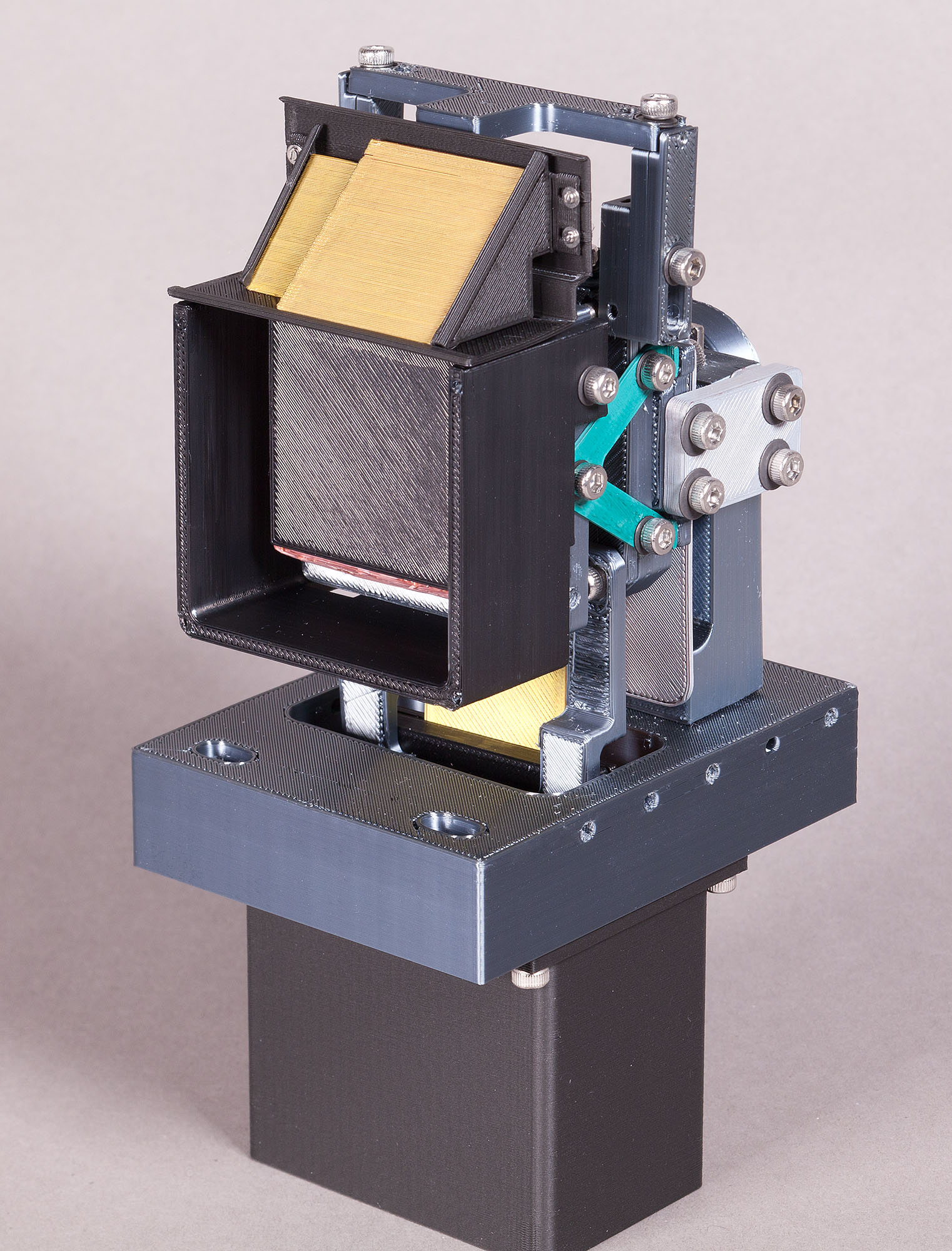}{0.45\textwidth}{(b): 3D print}}
    \caption{(a): Rendering of the Liger imager detector and pick-off mirror mounting and focus stage with all exterior baffling removed. The directional axes used throughout this paper are shown on the left hand side. The y-direction (green) is used to denote the height above the optical plate; the z-direction (blue) represents the direction of the beam path; and the x-direction (red) is parallel to the optical plate and perpendicular to the beam. The pick-off mirrors (gold) for both the lenslet and slicer IFS are made from one piece of Zerodur and are attached to the detector mounting so the two mirrors are coupled in position. A baffle snout is included around three sides of the detector which extends $\rm12\,mm$ in front of the detector face to block scattered light. The pick-off baffle lowers onto the top edge of this baffle snout and also extends to the same distance. The detector ASIC is held below the detector mount connected via a flexible cable. Adjustability is built into the mounting assembly in the focus position as well as the tip-tilt of the detector and pick-off mirrors. The height (y-direction) and x-direction adjustment will be made at the base of the mount. (b): A full scale 3D printed model of the same assembly with the baffling included around the ASIC. This model was built to test the planned assembly procedure.}
    \label{fig:rendering}
\end{figure}

The imager detector and the pick-off mirrors for the slicer and lenselt IFS are coupled to the same mounting assembly on the imager optical plate (see Figure \ref{fig:rendering}). The pick-off mirrors for the two spectrograph channels will be made from a single piece of Zerodur. The orientation of the detector and pick-offs will be semi-fixed to each other; the relative z-offset (into the beam) and rotation will be independently adjustable within a small range. The alignment of the assembly in the x-direction (parallel to the optical plate and perpendicular to the beam) and y-direction (height) as well as tip-tilt for both the detector and pick-off mirrors, are adjusted manually where the assembly mounts to the optical plate. An additional tip-tilt adjustment for the detector is built into the assembly, as well as a rotation adjustment for the pick-off mirrors. The optimal z-position will be determined by moving the assembly through a range of focus positions during alignment via a piezoelectric linear actuator. This will yield the optimal offset between the detector and pick-off mirrors, that then can be adjusted via set screws. Once optimal focus is determined between the detector and pick-off mirrors the assembly is locked.

\subsection{Focus Alignment}\label{sec:focus}
\begin{figure}[h]
    \centering
    \gridline{\fig{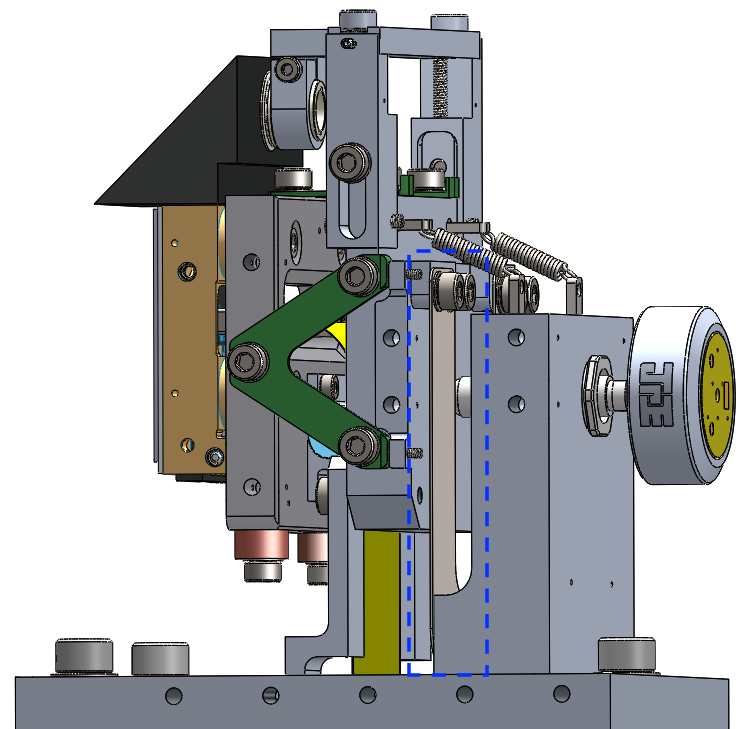}{0.4\textwidth}{(a): neutral position}
            \fig{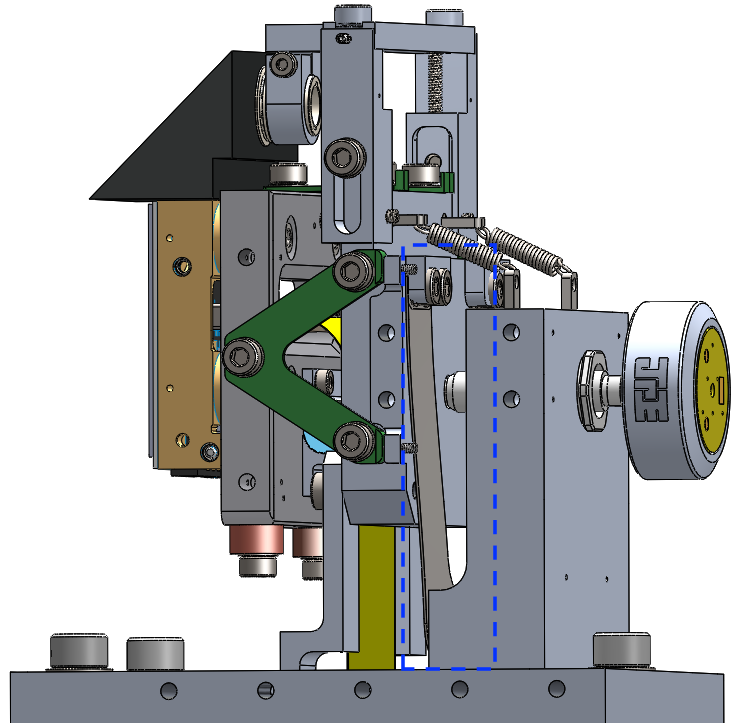}{0.4\textwidth}{(b): extended position}}
    \caption{Side view of the detector and pick-off mirror assembly showing the AISI 304 flexure (outlined in blue) component connecting the base with the rest of the mount as well as the extension springs which maintain planarity of the system throughout the range of possible focus positions. The image on the left shows the neutral position while the right image shows the flexure and mount at the end of the $\rm3\,mm$ travel range. During operation support brackets will be included on each side (loosely fastened) to further maintain planarity and lock the system at the optimal focus position. The brackets are excluded from these models to show the operation of the flexure.}
    \label{fig:focus_positions}
\end{figure}

The mounting assembly is designed to allow the detector and pick-off mirrors to move through a $\rm3\,mm$ range of possible focus positions during the alignment process. This is accomplished with the inclusion of a flexure between the mounting assembly base and the mounting points for the detector and pick-off mirrors. The flexure is made of AISI 304 stainless steel sheet metal cut into a ``U" shape that is bolted to the fixed base at the bottom and the mobile mount at the tops. As the piezoelectric linear actuator pushes on the mount, the arms of the flexure bend and extend forward as shown in Figure \ref{fig:focus_positions}. To maintain planarity of the detector face throughout the travel range, two extension springs are located above the arms of the flexure connecting the fixed base to the mobile  mount. The force from these springs pulls back on the top of the mount, preventing the detector face from tilting forward as the flexure bends. Two support brackets are included between the fixed base and the mobile mount. These supports have clearance slots that mate to threaded holes on the fixed base. The screws at this location will be kept loose during focus adjustment, after which the support brackets are secured to lock the assembly into place.

Both the detector and pick-off mirrors are coupled to each other as they are moved through the range of potential focus positions. There may be an offset between these best focus positions, in which case the mounting arm attaching the pick-off mirrors to the main assembly may be adjusted by $\rm \pm0.5\,mm$ in the z-direction with an accuracy of $\rm 100\,\mu m$.

\begin{figure}[h]
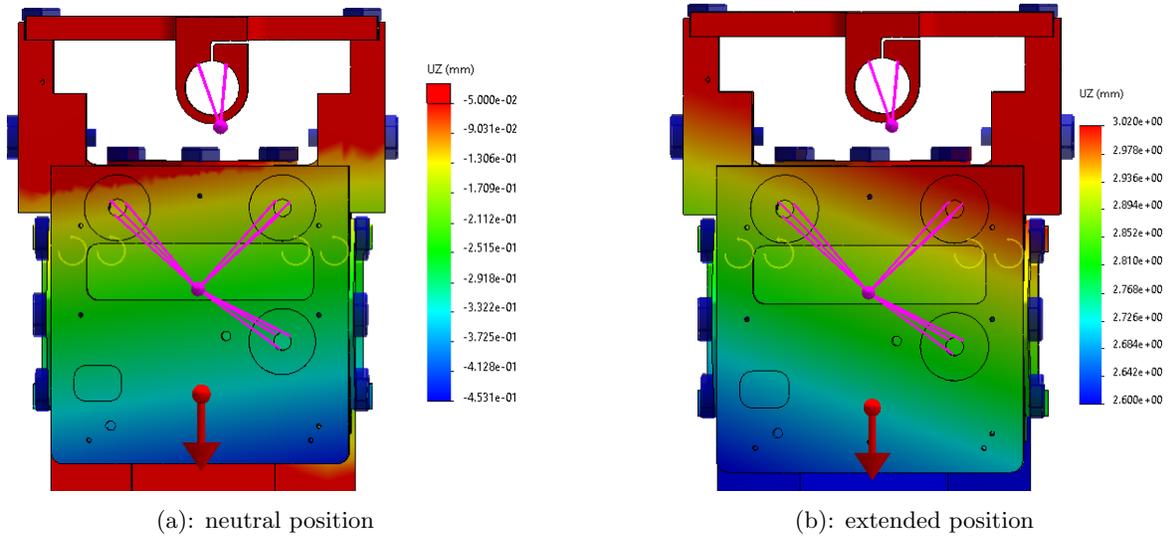

    \centering
    \gridline{\fig{pre_load_displacement.png}{0.4\textwidth}{(a): neutral position}
            \fig{20N_displacement.png}{0.4\textwidth}{(b): extended position}}
    \caption{Displacement from SolidWorks Simulation of static loading on detector stage at the two extremes of the focus range from $\rm0\,mm$ (a) to $\rm3\,mm$ (b). The color scales are adjusted to highlight the small deviation from planarity across the face of the front plate where the detector will mount at the three clearance holes. In (a), there is a $\rm\sim0.4\,mm$ difference across the face of the plate, resulting in an angle of $\rm0.46^\circ$. At the end of the focus range shown in (b), there is a $\rm\sim0.42\,mm$ difference across the face of the plate, resulting in an angle of $\rm0.48^\circ$. The change in the angle between these two extremes of the focus range is only $\rm 0.02^\circ$, much less than the tolerance in this dimension.}
    \label{fig:planarity_sims}
\end{figure}

It is important that there is no significant change in the tilt of the detector and pick-off mirrors while determining the optimal focus position. To check the planarity of the mounting block (and by extension the detector and pick-off mirrors) a simplified model is used in a static load simulation within the SolidWorks Simulation suite. The weight of the mounting assembly as well as remote masses for the detector and pick-off mirrors are included in the simulation as well as all bolts (with pre-tension) and spring parameters used in the design. The bottom of the mounting plate is designated as a fixture within the simulation. First, a baseline is determined by conducting a simulation with no force from the actuator to yield the displacement across the detector face under static load. Next a load is added at the location of the linear actuator to cause forward motion of the detector mounting. The two cases must result in a difference in the tip-tilt angles $\rm <0.2^\circ$ in order to meet the required tolerance. The displacement across the detector face mounting plate in both simulations are shown in Figure \ref{fig:planarity_sims}. As can be seen, there is a small change in the tilt of the detector mounting plane after the full $\rm3\,mm$ of linear motion. However, this change amounts to only $\rm 0.02^\circ$, significantly less than the $\rm 0.2^\circ$ tolerance.

\subsection{Positioning Adjustment}\label{sec:adjustment}
There are multiple adjustment points included in the detector assembly design which will allow fine-tuning during alignment. There are six degrees of freedom to the position of the pick-off mirrors and detector, although some have a limited range and/or require the use of shims at the base of the mount. Some of these adjustments cause position changes to both the detector and pick-off mirrors while others will only impact one. The separate adjustment of the detector and pick-off mirrors is particularly useful in cases where an offset is required (e.g. focus position) or when a tighter tolerance is required for one component than the other (e.g., tip-tilt). 

\begin{figure}[h]
    \centering
    \gridline{\fig{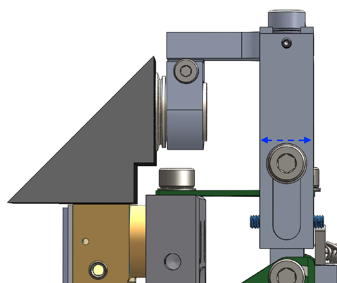}{0.4\textwidth}{(a): focus position}
            \fig{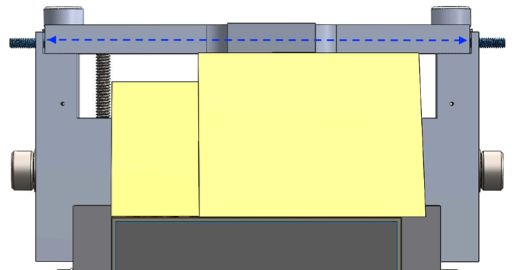}{0.4\textwidth}{(b): x-position}}
    \caption{Model views of where the pick-off mirror mounting attaches to the main assembly with baffling removed. Left: Side view; the blue arrow shows how the focus position of the pick-off mirrors can be adjusted independently of the detector using the highlighted set screws. This adjustment can be made over a range of $\rm\pm0.5\,mm$ with an accuracy of $\rm100\,\mu m$. Right: Front view; the blue arrow shows how the x-position of the pick-off mirrors can be adjusted independently of the detector over a range of $\rm\pm0.25\,mm$ with an accuracy of $\rm75\,\mu m$ using the highlighted set screws. The rotation of the pick-off mirrors can be independently adjusted at the mounting location of the pick-off frame.}
    \label{fig:pickoff_offset}
\end{figure}

First is the adjustment of the detector and pick-off distances in the z-direction (focus position). As discussed in the previous section, there is a flexure and actuator which allows both the detector and pick-off mirrors to be to be moved through the $\rm3\,mm$ focus range to find the optimal position. We have designed flexibility in the alignment procedure if there is an offset in the optimal position for the detector and the pick-off mirrors. If this occurs there is $\rm\pm0.5\,mm$ over which the pick-off mirror focus position can be adjusted independently of the detector with an accuracy of $\rm100\,\mu m$ (see Figure \ref{fig:pickoff_offset}a).

The height of the detector assembly may be adjusted by including shims between the optical plate and the mount. The height of the pick-off mirrors may be further raised above the detector in increments of as little as $\rm100\,\mu m$ using a pair of set screws. The rotation of the pick-off mirrors about the z-axis may be adjusted independently of the detector with an accuracy of $\rm0.136^\circ$ by raising either side of the frame using these same set screws. The rotation of both components can be changed by inclusion of shims on one side of the mounting plate. The position of both the detector and pick-off mirrors in the x-direction can be adjusted by pushing the assembly along clearance slots at base of the mount shown in Figure \ref{fig:rendering}. The pick-off mirror position can be independently adjusted in the x-direction by $\rm\pm0.25\,mm$ with an accuracy of $\rm75\,\mu m$ using the set screws on the side of the pick-off mounting frame (Figure \ref{fig:pickoff_offset}b).

\begin{figure}[h]
    \centering
    \gridline{\fig{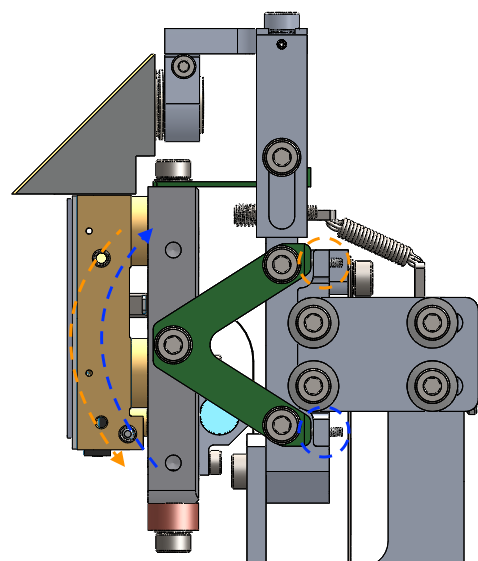}{0.38\textwidth}{(a): tip}
            \fig{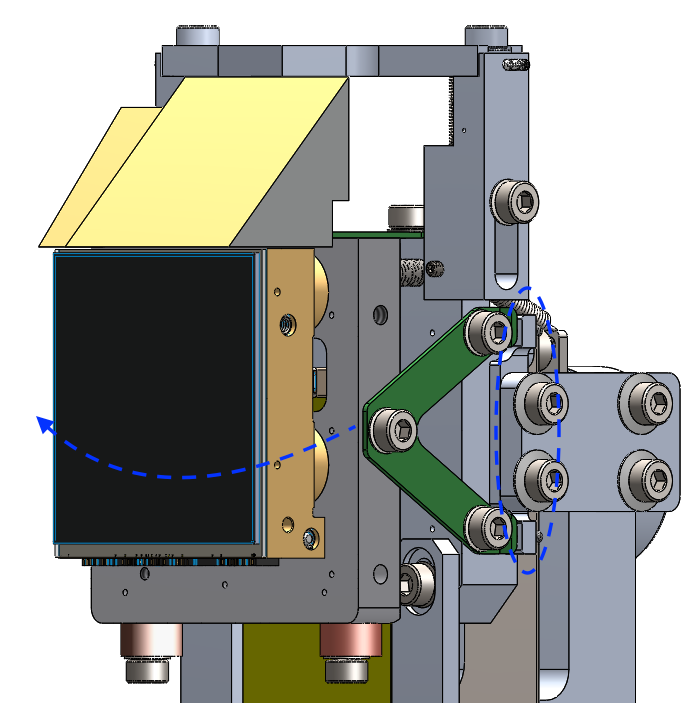}{0.4\textwidth}{(b): tilt}}
    \caption{Model view of the detector mounting assembly illustrating the adjustability in tip (left) and tilt (right) via set screws at the A frame feet. Left: pushing the top foot of the A frames on the sides will result in changing the tip in the direction of the orange arrow. Likewise, pushing the bottom foot will cause a change in direction following the blue arrow. The A frame on the other side of the detector should be pushed in the same way for adjustments to the tip. Right: pushing both the top and bottom feet of the A frame shown forward by the same amount will cause an adjustment to the tilt of the detector following the blue arrow. Conversely, pushing the top and bottom feet of the A frame on the opposite side of the detector will cause a tilt in the opposite direction.}
    \label{fig:TT_adjustability}
\end{figure}

The tip-tilt angle may be adjusted independently for the detector and pick-off mirrors. Set screws can push the feet of the A frames which hold the detector to the main mount, allowing for a range of $\rm\pm1.13^\circ$ of adjustability in tip and  $\rm\pm0.57^\circ$ in tilt (Figure \ref{fig:TT_adjustability}). The $\rm0.25^\circ$ tolerance is met in both dimensions, with a quarter turn of the set screw giving a $\rm0.17^\circ$ change in tilt and a $\rm0.09^\circ$ change in tip. The tilt of the pick-off mirrors can be adjusted using the same set screws shown in Figure \ref{fig:pickoff_offset}a that are used to set the focus offset. By moving one side of the pick-off frame to a closer or further offset position, slight adjustments to the tilt of the mirrors can be made with an accuracy of $\rm0.13^\circ$. To adjust the tip angle of just the pick-off mirrors, the tip of both the mirrors and detector must be adjusted via shims between the mounting plates, while the detector can then be separately adjusted at the A-frames to compensate. The accuracy for the adjustment of the tip-tilt angle is less than the tolerance ($\rm 0.2^\circ$ for the pick-off mirrors and $\rm 0.25^\circ$ for the detector) even when including the potential change in the tip angle at different focus positions.

\subsection{Baffling}\label{sec:baffling}
\begin{figure}[h]
    \centering
    \gridline{\fig{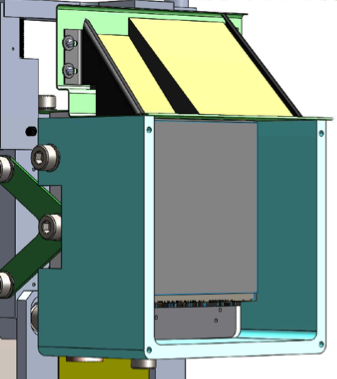}{0.4\textwidth}{(a): internal baffling}
            \fig{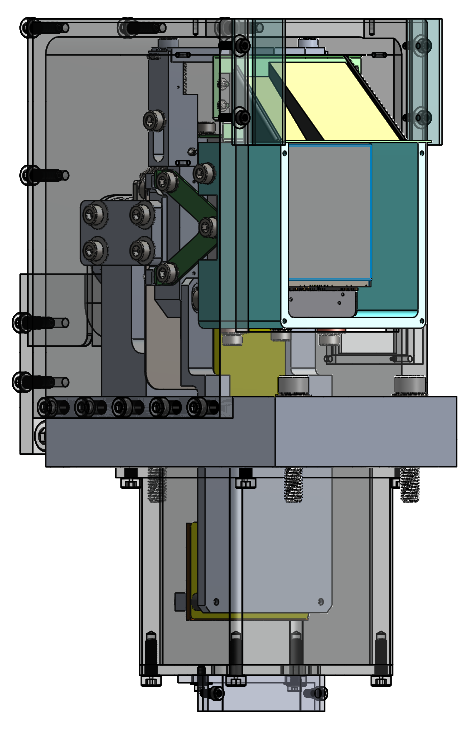}{0.35\textwidth}{(b): external baffling}}
    \caption{Model view showing the baffling around the detector and pick-off mirrors. Left: The three sided baffle around the detector (light blue) extends $\rm\sim0.7\,mm$ above its edge ($\rm<0.1\,mm$ above the Teledyne detector package) in order to protect this critical component while the pick-off mirrors are lowered into position. The sheet metal baffle (green) attached to the pick-off frame along with the mirror rests on this top edge and provides a roof over the detector to prevent ghosting from the re-imaging optics located after the pick-off mirrors. The silver baffles on the sides fold over the edges of the pick-off mirrors to prevent reflections off the corners. All of these baffles will be painted black; the color shown in the model is for illustrative purposes only. Right: Baffling (shown as transparent black) is also included around the full assembly including a separate box around the ASIC which extends below the optical plate.}
    \label{fig:baffling}
\end{figure}

Baffling is included directly around the detector and pick-off mirrors as well as a baffle around the entire mounting assembly for further reduction of stray light. A three-sided baffle slides over the detector, extending $\rm12\,mm$ in front of it. The top of this baffle is open due to the small clearance between the detector and the pick-off mirrors. Here, a sheet metal baffle made from shim stock is included that is coupled to the pick-off mirror wedge and lowered with it onto the assembly. This baffle rests on the top corners of the detector baffle and extends $\rm12\,mm$ past the front of the detector to prevent light from scattering off the spectrograph re-imaging optics. A small lip is folded over at each end of this baffle sheet to prevent reflections from the top and bottom corners of the pick-off mirror wedge. Two additional baffles made of shim stock will slide in from the sides with a fold to prevent reflections from the pick-off mirror edges. This internal baffling setup is shown in Figure \ref{fig:baffling}a, and the external baffling demonstrated in Figure \ref{fig:baffling}b.

\section{STRUCTURAL ANALYSIS}\label{sec:structural_analysis}
In order to determine how the assembly will respond to loading, a simplified model with only the mounting components is analyzed within the SOLIDWORKS Simulation suite. The baffling, detector, pick-off mirrors, and all fasteners are removed in this simplified model. The weight of the detector and pick-off mirrors are accounted for in the form of remote masses, and bolted joints are specified in the simulation with the fastener dimensions and pre-load. Like the simulation at each end of the focus range, the bottom of the mounting plate is designated as a fixture.

\begin{figure}[h]
    \centering
    \gridline{\fig{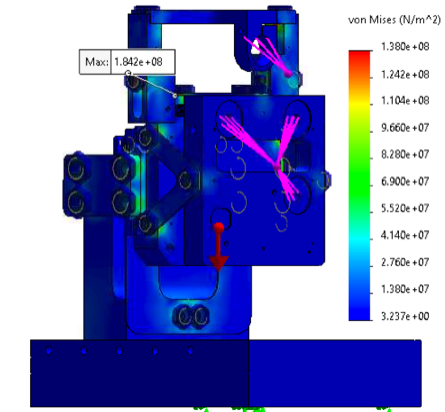}{0.4\textwidth}{}
            \fig{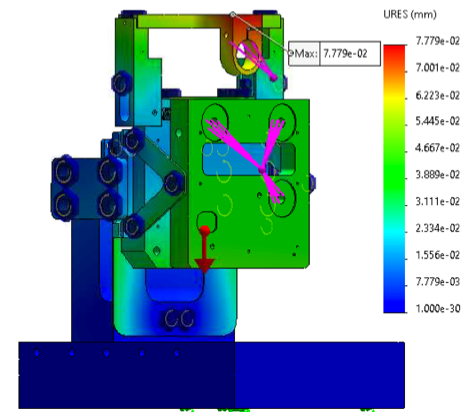}{0.43\textwidth}{}}
    \caption{Stress (left) and deflection (right) for simulation of the detector mounting under both static and an additional 4g shock load in the vertical direction (5g total). The maximum stress determined is $\rm <2/3\,S_y$ underneath bolted connections and $\rm <1/2\,S_y$ elsewhere with negligible deflections.}
    \label{fig:4g_load}
\end{figure}

With only the static load due to gravity the maximum stress is $\rm <2/3$ the yield strength of aluminum ($\rm S_y = 276\,MPa$). The areas of the highest stress are located under bolted connections due to pre-loading while elsewhere the stress is $\rm <1/2\,S_y$. The assembly is required to withstand an additional 4g's of loading during shipping and/or earthquakes. With this load applied in the vertical direction (adding to the static load due to gravity) the maximum stress in the model is still $\rm <2/3\,S_y$, located underneath the bolted connections at the A frames (Figure \ref{fig:4g_load}). This stress is also present here with only the static load due to gravity and is therefore likely due to compression from the bolt pre-load. Since the 4g load may be experienced during shipping it may occur along any direction.  With the shock load applied in the x- and z-directions the maximum stress is located under an A frame bolted connection instead, but is still $\rm <2/3\,S_y$. In all loading cases the stress outside of the bolted connections is $\rm <1/2\,S_y$.

If the best focus position involves full extension of the flexure, there will be a higher stress in the steel flexure itself. As shown in Figure \ref{fig:4g_load}, the mount as a whole will not fail since the support brackets will hold it up, but we also want to ensure that the flexure is not damaged so that focus adjustments can be made later in the life of the instrument. To investigate the worst case scenario with the highest stress, the 4g shock load is simulated with a deformed flexure in the most extended configuration (with the optimal focus furthest from the starting estimate).  This results in the stress throughout the flexure being well below the ultimate strength of AISI 304 steel (505MPa) except for at a single edge of the flexure. This is at the location of a modification to the deformed flexure model needed in order to create parallel surfaces for mating parts. This is likely creating an artificial stress concentration which will not be present in the fabricated part. Outside of this modified edge of the flexure, the stress is below the ultimate strength throughout, and outside of the compressive stress under fasteners, the stress is also below yield.

A frequency analysis was run in SOLIDWORKS Simulation using the simplified model in order to estimate the resonant frequencies. The mass of the detector and pick-off mirrors are again added as a remote load offset from the mounting point with the force of gravity included. The simulation was run searching for the first five frequency modes for the simplified model used in the load simulations as well as with the external baffling included (Table \ref{tab:modes_nobaffle} \& \ref{tab:modes_baffle} respectively). As can be seen from the lower mass participation factors in Table \ref{tab:modes_baffle}, the external baffling largely does not participate in the modal response.

\begin{table}
\centering
\caption{Frequencies and mass participation for the first five modes of the imager detector and pick-off mounting without external baffling \label{tab:modes_nobaffle}}
\begin{tabular}{|c|c|c|c|c|}
\hline
Mode & Frequency (Hz) & \multicolumn{3}{c|}{Mass Participation (\%)} \\
 &  & X-direction & Y-direction & Z-direction \\
 \hline \hline
1 & 345 & 0.2 & 28 & 14 \\
2 & 368 & 40 & 0.2 & 0 \\
3 & 480 & 0 & 0 & 2 \\
4 & 787 & 0 & 9 & 32 \\
5 & 804 & 0 & 0.3 & 0.8 \\
\hline
\end{tabular}
\end{table}

\begin{table}
\centering
\caption{Frequencies and mass participation for the first five modes of the imager detector and pick-off mounting with external baffling \label{tab:modes_baffle}}
\begin{tabular}{|c|c|c|c|c|}
\hline
Mode & Frequency (Hz) & \multicolumn{3}{c|}{Mass Participation (\%)} \\
 &  & X-direction & Y-direction & Z-direction \\
 \hline \hline
1 & 331 & 0.1 & 20 & 8 \\
2 & 355 & 27 & 0.1 & 0 \\
3 & 483 & 0 & 0 & 0.8 \\
4 & 744 & 0 & 5 & 23 \\
5 & 850 & 0.8 & 0 & 0 \\
\hline
\end{tabular}
\end{table}

The instrument requirements list frequencies that can be expected in the range of $\rm10 - 40\,Hz$ during shipping and $\rm8-80\,Hz$ during operation, so the resonant frequency of the mechanism should be at least $\rm >80\,Hz$. The frequencies determined in the SolidWorks simulations lie well outside of this range.

\section{SUMMARY}\label{sec:summary}
A single mounting assembly and housing will be used for the Liger imager detector and the IFS pick-off mirrors. This assembly will hold these optics $\rm<1\,mm$ away from each other while allowing adjustability in multiple axes.
The detector and pick-off mirrors can be adjusted both together and individually, although with a smaller range of individual adjustment. Baffling is included around the individual optics as well as the full assembly to protect from scattered light and ghosting off the surfaces of other optics.

Analysis was carried out on the imager detector and IFS pick-off mirror mounting assembly to verify that the design meets the requirements. The planarity of the mounting points were evaluated throughout the range of focus positions by simulating the mechanism reaction to a force at the location of the linear actuator. The deviation from planarity across the $\rm3\,mm$ range of travel was $\rm0.02^\circ$; well within the $\rm0.25^\circ$ tolerance. Next the strength of the mount was evaluated under both static and an additional 4g shock load. In both cases the maximum stress determined in the simulations is under bolt heads and is not expected to lead to failure. Modal frequencies were also determined using the SOLIDWORKS simulation tools, with all frequencies falling well above the $\rm8-80\,Hz$ range the system is expected to be subjected to.

The detector and pick-off mirror mount will be assembled and alignment performed in a custom cryogenic vacuum chamber designed for use with the Liger imager which will operate at a temperature below $\rm77 \, K$ and vacuum pressure of $\rm 10^{-5} \, Torr$\cite{Wiley2020}. The unique design of Liger allows simultaneous imaging and spectroscopy which both improves the performance of the instrument and provides useful science benefits. For example, simultaneous imaging in crowded field observations (e.g., globular clusters or the galactic center) can provide accurate astrometry and real time measurements of the telescope and instrument point-spread function. Having the light for the spectrograph modes first pass through the imager provides benefits such as improved background masking at the larger pupil located in the imager\cite{Cosens2020}. It also allows for improved AO correction in the imager without sacrificing the IFS performance. The design of the detector and IFS pick-off mirrors is critical to maintaining this performance. The closer the pick-off mirrors are to the detector, the lower the wavefront error is for both as it is best in the center and degrades with increasing radius.  

\acknowledgments 
 
The Liger instrumentation program is supported by the Heising-Simons Foundation, the Gordon and Betty Moore Foundation, University of California Observatories, and W. M. Keck Observatory.  

\bibliography{detector} 
\bibliographystyle{spiebib} 

\end{document}